\documentclass[prl,twocolumn,superscriptaddress,showpacs]{revtex4}
\usepackage{epsfig}
\usepackage{amssymb}
\usepackage{multirow}

\begin{document} 

\title{Van der Waals density functional: an appropriate exchange functional}

\author{Valentino R. Cooper}
\email{coopervr@ornl.gov} 
\affiliation{Materials Science and Technology Division, Oak Ridge, TN 37830-6114, USA}
\date{\today}
\begin{abstract}
In this paper, an exchange functional which is compatible with the
non-local Rutgers-Chalmers correlation functional (vdW-DF) is
presented.  This functional, when employed with vdW-DF, demonstrates
remarkable improvements on intermolecular separation distances while
further improving the accuracy of vdW-DF interaction energies.  The
key to the success of this three parameter functional is its reduction
of short range exchange repulsion through matching to the gradient
expansion approximation in the slowly varying/high density limit while
recovering the large reduced gradient, $s$, limit set in the revised
PBE exchange functional.  This augmented exchange functional could be
a solution to long-standing issues of vdW-DF lending to further
applicability of density functional theory to the study of relatively
large, dispersion bound (van der Waals) complexes.
\end{abstract}


\maketitle 

Van der Waals, or London dispersion, interactions have profound
importance in bio-organic systems as well as many novel materials
being investigated for energy applications.  
Despite the importance of these systems and applications, first
principles simulations have been greatly lacking.  The primary reason
for this is the inability of traditional density functional theory
exchange-correlation functionals to account for long-ranged, van der
Waals interactions.  This has limited first-principles investigations
to quantum chemical methods which, due to their computational expense,
are only capable of modeling fragments of the true material; thus
often overlooking some of the more salient features of these systems.

\begin{figure}
\includegraphics[width=2.5in]{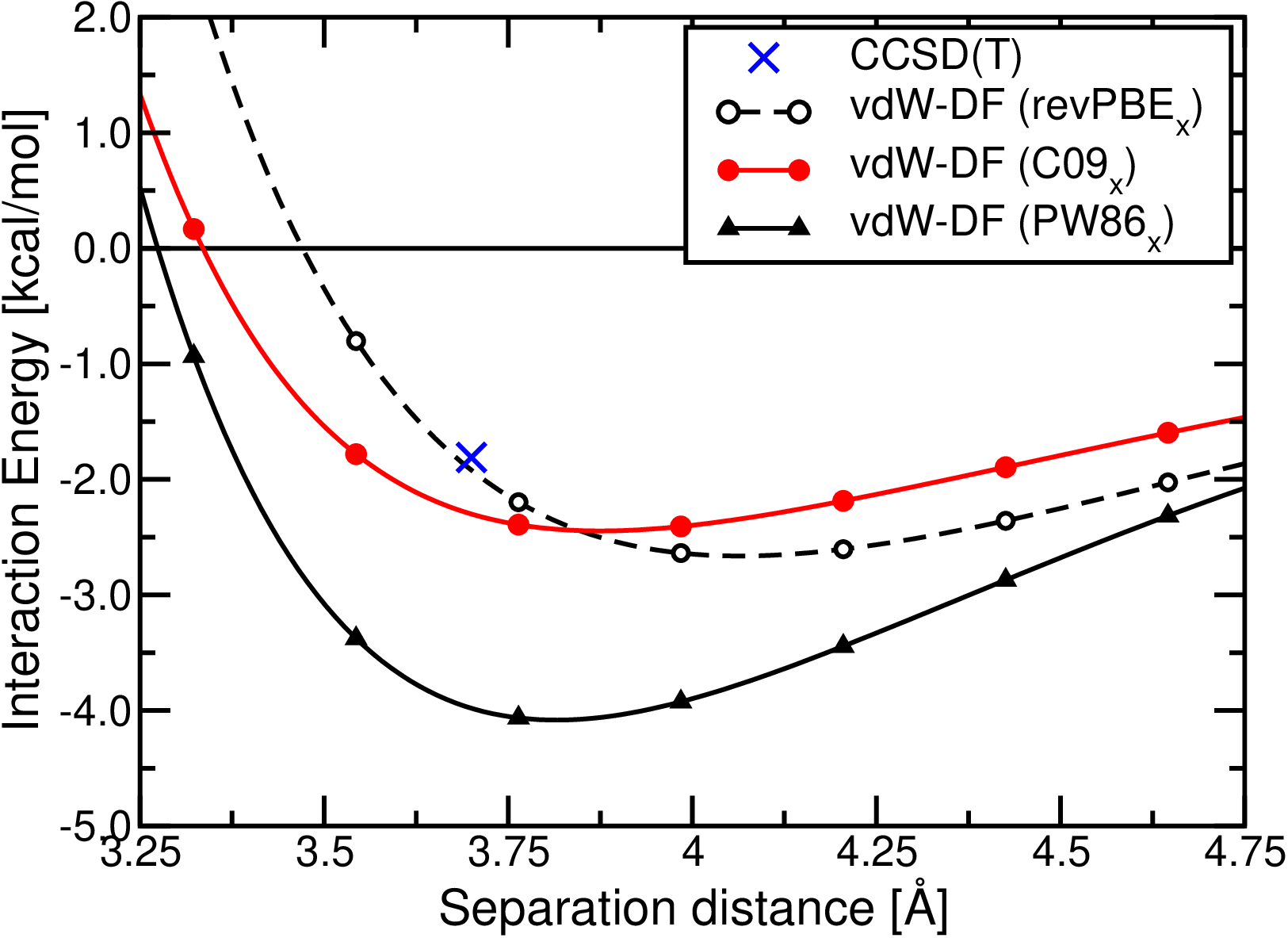}
\caption{\label{benz_fig} (color online) Interaction energy as a
function of separation distance for the benzene dimer in the stacked
sandwich configuration.  The vdW-DF results with revPBE, CO9 and PW86
for exchange are represented by a black dashed line with open circles,
a red solid line with closed circles and a solid blue line with closed
triangles, respectively.  Benchmark CCSD(T) data~\cite{Tsuzuki02p104}
(blue cross) are plotted for reference.}
\end{figure}

A promising solution to the problem of balancing speed and scalability
with accuracy lies in the non-local correlation functional of the
Rutgers-Chalmers collaboration, the aptly named van der Waals density
functional (vdW-DF).\cite{Dion04p246401,Thonhauser07p125112} This
method includes long-range dispersion effects as a simple perturbation
to the local density approximation correlation term and has been
extremely successful in describing a diverse group of materials
properties - ranging from molecules to bulk polymers and the
adsorption of molecules to surfaces and within bulk
materials.\cite{Langreth09p084203} Recent developments show that
self-consistency gives no appreciable differences in computed
interaction energies\cite{Dion04p246401,Thonhauser07p125112} and such
a non-local functional can in fact be incorporated in an extremely
efficient manner.\cite{Roman-Perez09p0812.0244} However, the
overwhelming success of vdW-DF is marred by its consistent
overestimation of intermolecular distances.\cite{Langreth09p084203}
Analysis of various generalized gradient approximation exchange
functionals (GGA$_{\rm x}$) indicate that traditional functionals are
either too repulsive at short distances or incorrectly exhibit some
``correlation'' binding at larger distances.  The standard functional
used within the vdW-DF, the revised Perdew-Burke-Erzenhoff functional
(revPBE),~\cite{Zhang98p890} unfortunately gives too much repulsion at
short distances.  Replacing revPBE with Hartree-Fock (HF) exchange
shows improvements in the inter-species separation distances obtained
with vdW-DF but at the cost of over-binding, i.e.  considerably larger
interaction energies than obtained via
CCSD(T).\cite{Puzder06p164105,Vydrov08p014106} Recent work suggests
that for many dispersion bound materials the PW86
functional~\cite{Perdew86p8800} most closely matches HF
exchange.\cite{Kannemann09p719,Murray09p} Similarly when applied with
vdW-DF it also strongly overbinds (see Fig.~\ref{benz_fig}).

In this paper, a GGA$_{\rm x}$ that may be more suitable for use with
the vdW-DF correlation functional is proposed.  This functional is
derived through the introduction of an enhancement factor which obeys
two specific constraints: (i) matching to the gradient expansion
approximation (GEA)~\cite{Sham71p} in the slowly varying/high density
limit and (ii) a smooth asymptote to the upper bound empirically set
in revPBE exchange.  Initial results indicate dramatic improvements in
vdW-DF separation distances while retaining the accuracy of this
method for a range of systems.  Most notable are improvements in the
interaction energies and the intermolecular/interplanar separation
distances obtained for S22 database structures\cite{Jurecka06p1985}
and graphite.

The general formula of a GGA$_{\rm x}$ can be written as:
\begin{equation}
E^{\rm GGA}_{\rm x} = \int{d^3r n \epsilon^{\rm unif}_{\rm x}(n) F_{\rm x}(s)},
\end{equation}

\noindent where $\epsilon^{\rm unif}_{\rm x}(n)$ is the exchange
energy per particle in a uniform gas ($\epsilon^{\rm unif}_{\rm x}(n)
= -3ek_F/4\pi$ with $k_F=[3\pi^2n]^{1/3}$) and $F_{\rm x}(s)$ is the
enhancement factor which is a function of $s=\nabla n/(2k_Fn)$.  This
form of exchange ensures proper, uniform density
scaling~\cite{Levy85p2010} where $F_{\rm x}(s)=1$ simply gives LDA
exchange.  In general, the enhancement factor is chosen such that
$F_{\rm x}(0)=1$.  Here we design an $F_{\rm x}(s)$ to fulfill two
further criteria:

(i) To reduce the short range exchange repulsion, in the limit of
$s\rightarrow0$, i.e. for slowly varying/high densities, the
functional approaches the GEA:

\begin{equation}
F^{GEA}_x(s) = 1 + \mu s^2
\end{equation}

\begin{figure}
\includegraphics[width=2.5in]{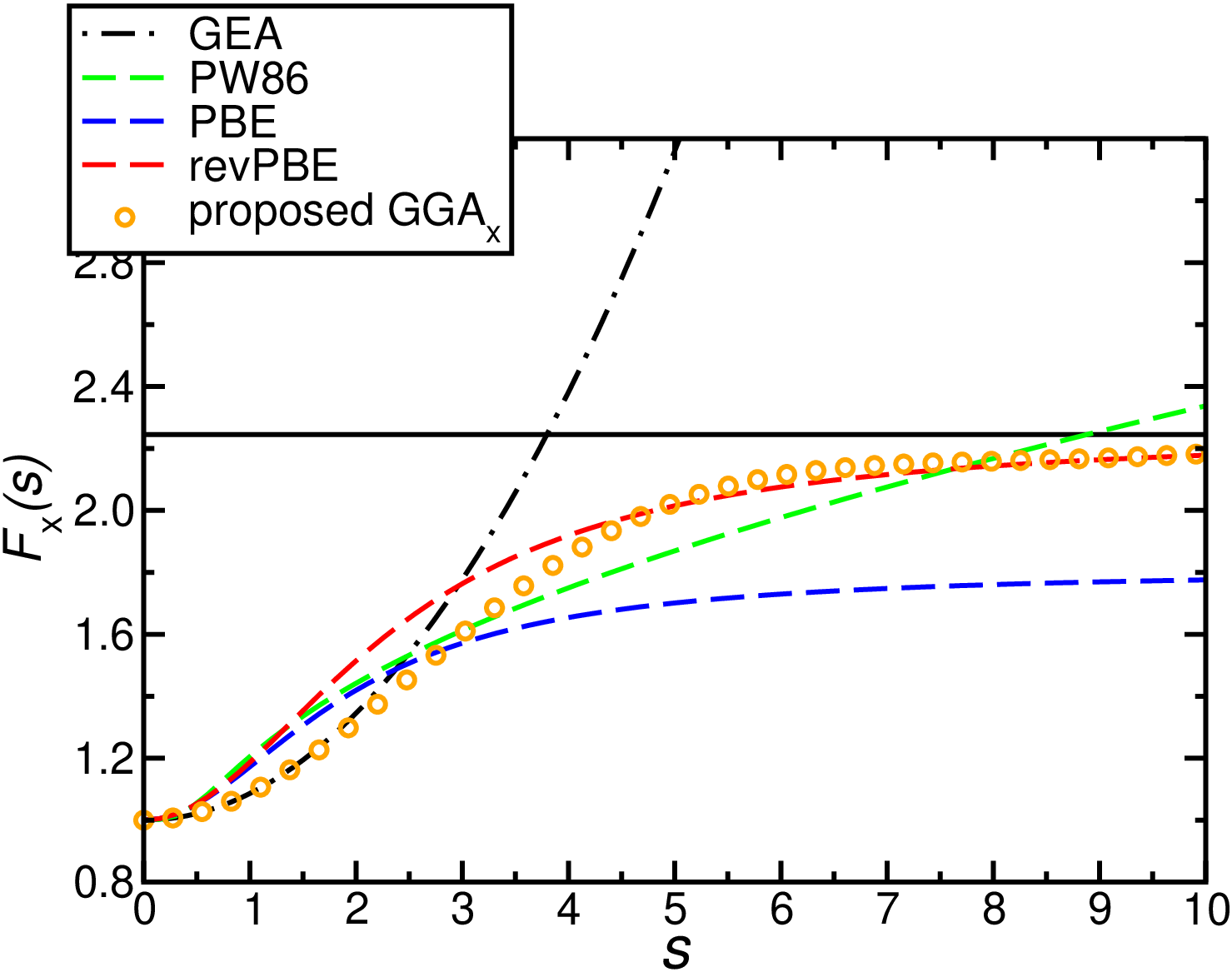}
\caption{\label{EnhFactfig}(color online)Enhancement factor for various GGA$_x$
functionals.  The blue, red and green dashed lines represent the
enhancement factor of PBE,\cite{Perdew96p3865}
revPBE~\cite{Zhang98p890} and PW86,\cite{Perdew86p8800} respectively.
The black, dashed-dot line is the GEA and the orange, open circles are
points from the proposed C09$_x$. The solid, black line indicates the
revPBE upper bound.}
\end{figure}

\noindent where $\mu$ = 0.0864.\cite{Sham71p} Figure~\ref{EnhFactfig}
depicts the enhancement factor of a number of GGA$_{\rm x}$.  It can
be clearly seen that many of these functionals deviate quite quickly
from the GEA.  Note that this constraint is similar to that used in
the recent PBEsol exchange functional which was designed to restore
the gradient expansion in order to remove artificial bias towards free
atoms.\cite{Perdew08p136406} Here, decreasing the enhancement factor
for small $s$ (thus restoring the GEA) leads to a reduction in the
short range repulsion in the GGA$_{\rm x}$.

(ii) The second constraint used in the proposed GGA$_{\rm x}$ is to
asymptote the revPBE upper bound of 2.245 in the large
$s$-limit.\cite{Zhang98p890} This simple bound is taken to be
compatible with previous applications of the revPBE exchange
functional with the non-local vdW-DF correlation term.  Here, we find
that an empirical $F_{\rm x}(s)$ bound similar to revPBE gives the
best interaction energies.

Using these constraints a simple, smooth, function can be constructed
in the form:

\begin{equation}
\label{EnhFacteqn}
F_{\rm x}(s) = 1 + \mu s^2 e^{-\alpha s^2} + \kappa (1-e^{-\alpha s^2/2})
\end{equation}

\noindent with $\mu$=0.0617, $\kappa$=1.245, $\alpha$=0.0483.
Fig.~\ref{EnhFactfig} displays the enhancement factor of
eqn.~\ref{EnhFacteqn} along with that for other GGAs.  The parameters
were determined by simultaneously fitting Eqn.~\ref{EnhFacteqn} to GEA
for $s<1.5$ and to revPBE for $s>8.0$.  This fitting domain was
arbitrarily chosen to allow for a decrease in $F_{\rm x}(s)$ for small
$s$ and a smooth recovery of revPBE for large values of $s$. (In
accordance with previous naming conventions this functional shall be
referred to as C09$_x$).  The complimentary exchange potential can be
easily constructed from the functional derivative of
Eqn.~\ref{EnhFacteqn} as shown in Eqn. 24 of
Ref.~\onlinecite{Perdew86p8800}.

To test the compatibility of the proposed C09$_x$ with vdW-DF,
self-consistent calculations within a modified version of the Abinit
plane wave code~\cite{Gonze02p478} were performed.  All calculations
were carried out with a 30 Ha planewave cutoff and a single k-point at
$\Gamma$. To reduce the effects of periodic images, all simulation
cells were padded with at least 10 \AA\ of vacuum in all directions.

\begin{table*}[ht]
\caption{\label{S22_table}Computed interaction energies for the S22
model data set.\cite{Jurecka06p1985} C09 values in parentheses are for
full geometry optimizations. For comparison, vdW-DF interactions
energies using the revPBE and PBE exchange functionals taken from
Ref.~\onlinecite{Gulans09p201105} (unless otherwise noted) are
listed. Deformation energies are not included.  Energies are in
kcal/mol.}
\begin{tabular}{lccccc}
\hline
\hline
\multirow{2}{*}{No.}& \multirow{2}{*}{Complex} & \multicolumn{3}{c}{vdW-DF}  & Benchmark\\
& &  revPBE\footnote{structures optimized for separation distance} & PBE\footnotemark[1] & C09 & CCSD(T)/CBS\\
\hline
\multicolumn{2}{l}{Hydrogen bonded complexes} \\
1 & (NH$_3$)$_2$ ($C_{\rm 2h}$) & 2.44 & 3.71 & 2.88 (2.92) & 3.17 \\ 
2 & (H$_2$O)$_2$ ($C_{\rm s}$) & 4.08 & 5.58 & 4.99 (5.15) & 5.02 \\
3 & Formic acid dimer ($C_{\rm 2h}$) & 14.07 & 18.28 & 20.12 (24.59)  & 18.61  \\
4 & Formamide dimer ($C_{\rm 2h}$)& 12.50 & 16.10 & 16.15 (17.15) & 15.96 \\
5 & Uracil dimer ($C_{\rm 2h}$) & 16.17 & 20.59 & 20.93 (22.36)& 20.65 \\
6 & 2-pyridoxine $\cdot$ 2-aminopyridine ($C_{\rm 1}$) & 14.02 & 17.94 & 17.67 (19.66) & 16.71 \\
7 & Adenine $\cdot$ thymine WC ($C_{\rm 1}$) & 15.19\footnote{Ref.~\onlinecite{Cooper08p1304}} & 17.57 & 17.25 (19.16) & 16.37  \\
\multicolumn{2}{l}{Complexes with predominant dispersion contribution}\\
8 & (CH$_4$)$_2$ ($D_{\rm 3d}$) & 0.88 & 1.55 & 0.51 (0.51) & 0.53 \\  
9 & (C$_2$H$_4$)$_2$ ($D_{\rm 3d}$) & 1.41 & 2.68 & 1.16 (1.16) & 1.51  \\
10 & Benzene $\cdot$ CH$_4$ ($C_{\rm 3}$)& 1.57\footnote{Ref.~\onlinecite{Hooper08p891}} & 2.51 & 1.50 (1.71) & 1.50 \\
11 & Benzene dimer ($C_{\rm 2h}$) & 2.74\footnote{Ref.~\onlinecite{Puzder06p164105}} & 4.96 & 3.32 (3.40) & 2.73  \\
12 & Pyrazine dimer ($C_{\rm s}$) & 3.87 & 6.25  & 4.74 (4.75) &  4.42 \\
13 & Uracil dimer ($C_{\rm 2}$) & 9.41\footnote{Ref.~\onlinecite{Li09p11166} } & 12.91 & 10.31 (10.41) & 10.12  \\
14 & Indole $\cdot$ benzene ($C_{\rm 1}$) & 4.34 & 6.25  & 5.44 (5.48) & 5.22 \\
15 & Adenine $\cdot$ thymine stack ($C_{\rm 1}$)& 10.60\footnotemark[5] & 14.74 & 12.79 (12.73) & 12.23 \\
\multicolumn{2}{l}{Mixed complexes}\\
16 & Ethene $\cdot$ ethine ($C_{\rm 2v)}$) & 1.55 & 2.38 & 1.61 (1.60) & 1.53  \\
17 & Benzene $\cdot$ H$_2$O ($C_{\rm s}$) & 2.72\footnote{Ref.~\onlinecite{Li08p9031}} & 4.15 & 3.25 (3.24) & 3.28 \\
18 & Benzene $\cdot$ NH$_3$ ($C_{\rm s}$) & 1.87 & 3.18 & 2.28 (2.27) & 2.35 \\
19 & Benzene $\cdot$ HCN ($C_{\rm s}$)& 3.87 & 5.488 & 4.51 (4.48) & 4.46 \\
20 & Benzene dimer ($C_{\rm 2v)}$) & 2.05\footnotemark[4] & 3.98 & 2.85 (2.84) & 2.74 \\
21 & Indole $\cdot$  benzene T-shape ($C_{\rm 1}$) & 4.72 & 6.90 & 5.75 (5.71) & 5.73 \\
22 & Phenol dimer ($C_{\rm 1}$)& 5.81 & 8.51 & 7.00 (7.20) & 7.05 \\
\hline
\multicolumn{2}{l}{Avg. \% deviation}& 18 & 36 & 5 (9) & - - -\\
\hline 
\hline
\end{tabular}
\end{table*}

The interaction energy, $\Delta E^{\rm int}$, as a function of
separation distance, $d_{\rm sep}$, for the benzene dimer stacked in
the sandwich configuration is plotted in Fig.~\ref{benz_fig}. A
comparison of vdW-DF with the standard revPBE exchange functional
(vdW-DF$^{\rm revPBE}$) and the exchange functional (vdW-DF$^{\rm
C09_x}$) using the enhancement factor of eqn.~\ref{EnhFacteqn} shows a
significant shortening of the separation distance from 4.07~\AA\ to
3.87~\AA.  The vdW-DF$^{\rm C09_x}$ is now in much better agreement,
with regards to both interaction energy and separation distance, with
both the benchmark CCSD(T) ($\Delta E^{\rm int}$ = 1.81 kcal/mol and
$d_{\rm sep}$ = 3.70~\AA) as well as SAPT(DFT) ($\Delta E^{\rm int}$ =
1.67 kcal/mol and $d_{\rm sep}$ = 3.80~\AA)~\cite{Podeszwa05p488}
calculations.  As previously mentioned, the PW86 exchange functional,
which was recently reported to mimic Hartree-Fock exchange for
dispersion bound complexes, gives excellent separation distances, but
significantly over estimates the interaction energy.

\begin{figure}[h]
\includegraphics[width=2.5in]{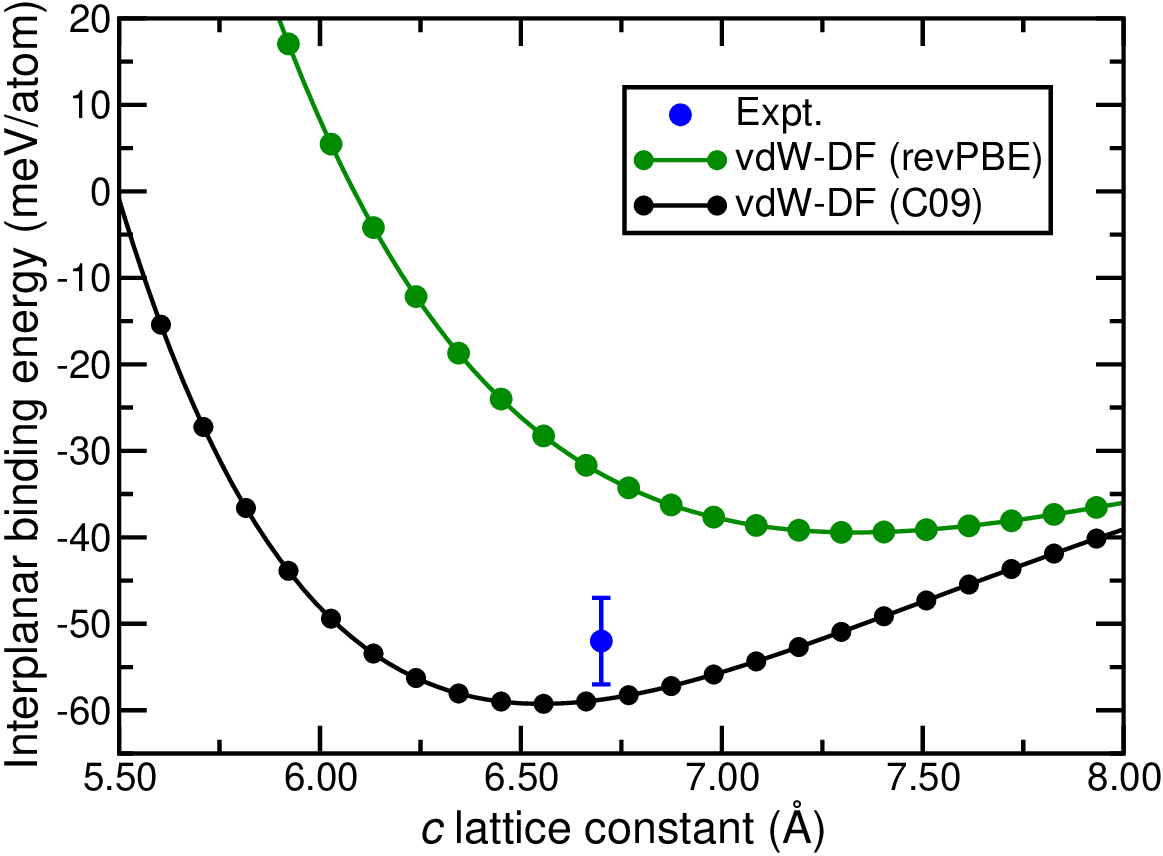}
\caption{\label{argon_fig} (color online) Interaction energy as a
function of interplanar separation distance for graphite.  The vdW-DF
results with revPBE and CO9 for exchange are represented by a black
dashed line with open circles and a solid black line with closed
triangles, respectively.  A recent experimental
value~\cite{Zacharia04p155406} (blue open circles with error bars) is
plotted for reference.}
\end{figure}

To further illustrate the value of the C09$_{\rm x}$, the interaction
energy as a function of $c$ lattice parameter for graphite is plotted
in Fig.~\ref{argon_fig}.  Here, C09$_{\rm x}$ offers significant
improvements in both the value of graphite lattice constants as well
the interplanar interaction energy.  Our computed lattice constant of
6.56 \AA\ is within 2\% of experiment (6.70
\AA).\cite{Zacharia04p155406} The interplanar binding energy (-59
kcal/mo)is also greatly improved (-52 $\pm$ 5 kcal/mol for
experiment\cite{Zacharia04p155406}).  Note, vdW-DF$^{\rm {revPBE_{\rm
x}}}$ gives interaction energies and lattice constants of -39 kcal/mol
and 7.35 \AA, respectively.

A more stringent evaluation of the accuracy of the functional can be
gained through comparison with the benchmark S22 database of Jurecka
et al.\cite{Jurecka06p1985} This database contains the interaction
energies and structures of 22 structures with varying degrees of
hydrogen bonding and vdW interactions computed with CCSD(T)
extrapolated to the complete basis set limit and is currently accepted
as the gold standard for theoretical methods used to study systems
with significant dispersion interactions.  Recently, Gulans and
coworkers examined the S22 database using vdW-DF with both the revPBE
and PBE exchange functionals.\cite{Gulans09p201105} In general, they
found reasonable agreement with the interaction energies of the S22
database; with PBE producing better results for hydrogen bonded
complexes and revPBE showing less deviation for dispersion dominated
interactions.  However, there results were all for vdW-DF optimized
structures; which always give too large separation distances.

Table~\ref{S22_table} lists the computed interaction energies for the
S22 database.  Here, it is evident that across the board vdW-DF$^{\rm
C09_x}$ is in much better agreement with the CCSD(T) benchmark values
than when vdW-DF is used with either the PBE or revPBE functionals.
In fact, vdW-DF$^{\rm C09_x}$ has an average percent deviation of 5\%
(9\% if full geometry optimizations were performed); far less than
revPBE and PBE (17\% and 36\%, respectively).  Even more important is
the fact that these interaction energies were obtained using the
published S22 geometries - without any adjustment of dimer separation
distances, demonstrating once again the improvement that this
functional offers with regards to both interaction energies and
determining optimum separation distances.

All 22 structures were subsequently relaxed such that the forces on
all the atoms were less than 0.02 eV/\AA.  These values are listed in
parentheses in the Table~\ref{S22_table}.  Analysis of the relaxed
geometries indicate that the majority of the deviations are related to
changes in internal bond lengths.  This is evident in the larger
changes in the vdW-DF$^{\rm C09_x}$ interaction energies for
hydrogen-bonded structures.

In summary, an exchange functional that is compatible with the
Rutgers-Chalmers van der Waals correlation functional is proposed.
This functional was derived to closely match the enhancement factor
$F(s)$ of the gradient exchange approximation for values of 0 $<$ s
$<$ 1.5, while having an asymptote to the revPBE bound of 2.245.  In
general, this functional shows significant improvements over the
previous revPBE exchange.  In particular, vdW-DF$^{\rm C09_x}$ offers
better agreement with the benchmark S22 database with an average
deviation of only 5\% at the intermolecular separation distances of
the published geometries.  This is a feat which far surpasses that of
previous vdW-DF calculations which required larger separation
distances. It should be noted that as previously pointed out that
while the restoration of the gradient expansion gives improved
dispersion interactions it is expected to worsen atomization
energies.\cite{Perdew08p136406} Nevertheless, these results highlight
the promise of this functional for use with the vdW-DF method and may
offer a pathway to even more accurate first-principles calculations of
dispersion bound systems.

\acknowledgements VRC would like to thank David C. Langreth for many
valuable discussions.  VRC also acknowledges comments from Kieron
Burke and discussions with Alaska Subedi. This work was supported by
U.S. Department of Energy, Division of Materials Sciences and
Engineering. This research used resources of the National Energy
Research Scientific Computing Center, which is supported by the Office
of Science of the U.S. Department of Energy under Contract
No. DE-AC02-05CH11231.

\end{document}